\documentclass[12pt]{article}
\usepackage{pic02}
\usepackage{hyperref}
\usepackage{url}
\usepackage{graphicx}
\begin{document}
\title{\bf $J/\psi$ Radiative Decays}
\author{
Xiaoyan SHEN\\
for the BES Collaboration\\[0.5cm]
{\em Institute of High Energy Physics}\\ 
{\em Chinese Academy of Sciences}\\
{\em Beijing, P. R. China}}
\maketitle

\begin{figure}[h]
\begin{center}
%

\vspace{4.5cm}
\end{center}
\end{figure}
\vspace{-3cm}
\baselineskip=14.5pt
\begin{abstract}
The previous results of $J/\psi$ radiative decays from MARKIII, DM2,
Crystal Ball and BESI are briefly reviewed in this talk. The main part of
this talk focuses on presenting the recent results from BESII 
$5.8 \times 10^7$ $J/\psi$ data, including the Partial Wave Analysis (PWA)
results, the measurement of $\eta_c$ mass, as well as search for some
interesting states.     
\end{abstract}

\baselineskip=17pt

\section{Introduction}
One of the distinctive features of QCD as a non-Abelian gauge theory is the
interaction of quarks and gluons, which predicts the existence
of other types of hadrons with explicit gluonic degrees of
freedom -- glueballs and hybrids. The indirect evidence for gluon-gluon
interactions has been obtained at high energies. However, glueballs,
the bound states of gluons, and hybrids predicted by QCD, have
not been confirmed yet. Therefore, the observation of glueballs and hybrids
is, to some extent, a direct test of QCD, and the study of the
glueball and hybrid spectroscopy, together with
the meson and baryon spectroscopy will be a good laboratory
for the study of the strong interactions in the strongly coupled
non-perturbative regime.

Many QCD-based models and calculations,
for example, bag models \cite{bag}, flux-tube models \cite{flux}, QCD sum
rules \cite{qcdsum} and lattice QCD \cite{lqcd}
are developed to make predictions to the properties of glueballs and
hybrids. Of them, lattice QCD is considered as the most relevant since
it originated from the first principle of QCD. 

According to the calculations of different lattice QCD groups\cite{lqcd},
the lightest glueball is found to be a $0^{++}$ state with the mass in
the region of 1.5-1.7 GeV, while the next lightest glueball is a $2^{++}$ 
with the mass around 2.3GeV.

\begin{figure}[htbp]
\centerline{\hbox{
\includegraphics[width=8cm]{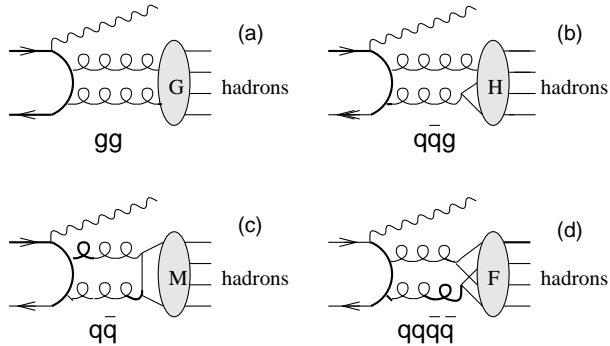}
       }}
 \caption{\it
     $J/\psi$ decays
    \label{jpsi-decays} }
\end{figure}

$J/\psi$ decay is a good lab. for the study of hadron spectroscopy,
as well as glueball and hybrid search. Fig.\ref{jpsi-decays} shows the
Feymann diagrams of some $J/\psi$ decays. A naive estimation of the
production rate of various particles based simply on counting powers of
the electromagnetic and strong coupling constants yields,\\

\[
\Gamma(J/\psi \to \gamma G) \sim O(\alpha \alpha_s^2),
~~\Gamma(J/\psi \to \gamma H) \sim O(\alpha \alpha_s^3)
\]
\[
\Gamma(J/\psi \to \gamma M) \sim O(\alpha \alpha_s^4),
~~\Gamma(J/\psi \to \gamma F) \sim O(\alpha \alpha_s^4)
\]

\noindent
where, G, H, M and F represent glueball, hybrid state, meson and 
four-quark state, respectively. Apparently, glueballs have enhanced 
production rates in $J/\psi$ radiative decays.

\section{Previous results on $J/\psi$ radiative decays}

Many experiments at the $e^+ e^-$ storage ring facilities,
such as Crystal Ball, MARKIII, DM2 and BES 
have been dedicating to the study of the hadron spectroscopy and
the search for non-$q \bar q$ states through $J/\psi$ radiative
decays.

DASP, MARKII, MARKIII, DM2, Crystal Ball and BES studied the
radiative production of $q \bar q$ mesons, such as tensors, scalars,
pseudoscalars and axialvectors, and measured the resonance 
parameters, decay branching ratios and polarization parameters.
In search for glueballs and new resonances, these experiments
provided much information on $f_0(1500)$, $f_0(1710)$, 
$\iota/\eta(1440)$ and $\xi(2230)$.
 
\section{Preliminary results from BESII $J/\psi$ data}

\begin{figure}[htbp]
  \centerline{\hbox{ \hspace{0.2cm}
    \includegraphics[width=5.5cm,height=5.0cm]{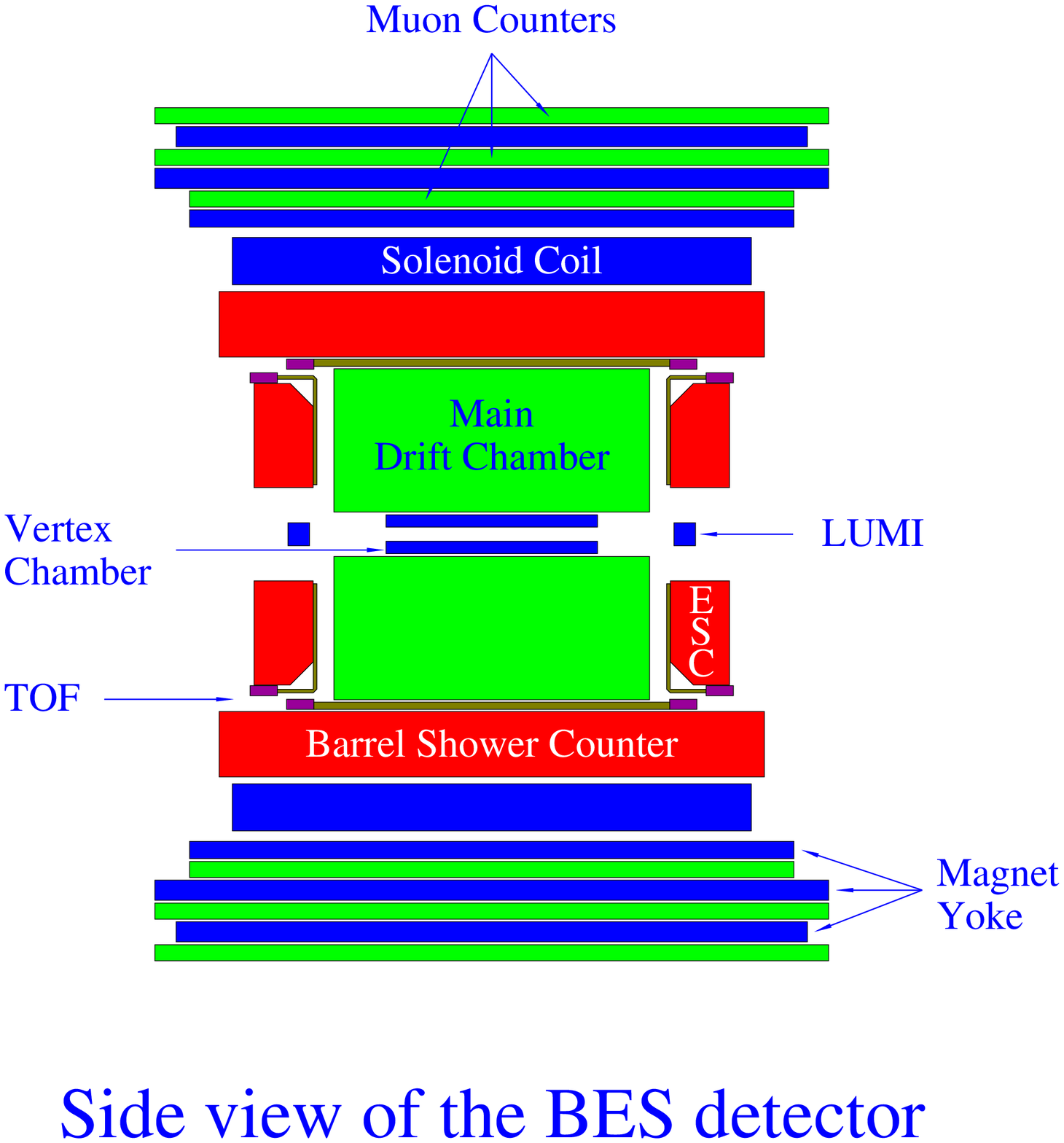}
    \hspace{0.3cm}
    \includegraphics[width=4.4cm,height=5.0cm]{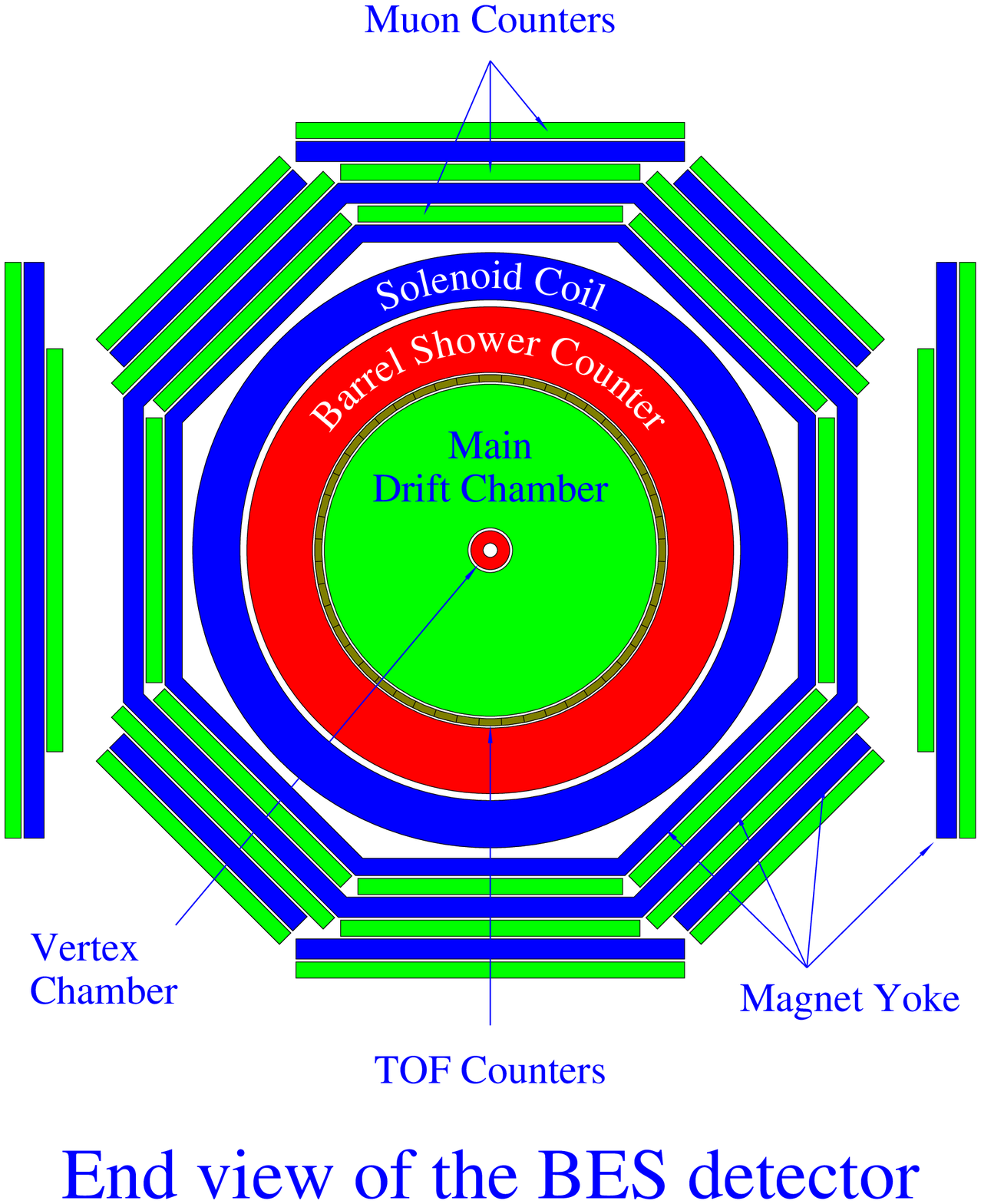}
    }
  }
 \caption{\it
      BES detector
    \label{bes} }
\end{figure}

BES, shown in Fig. \ref{bes}, is a large general purpose solenoidal detector 
at the Beijing Electron
Positron Collider (BEPC). The beam energy range is from 1.0 to 2.8 GeV and
the luminosity at $J/\psi$ peak is around $5 \times 10^{30} cm^{-2}s^{-1}$.
The details of BESI are described in ref. \cite{BES}.
\noindent
The upgrades of BESI to BESII\cite{besii} include the replacement of 
the central drift chamber with a vertex chamber composed of 12 tracking layers, the
installation of a new barrel time-of-flight counter (BTOF) with a
time resolution of 180ps and the installation of a new main
drift chamber (MDC), which has 10 tracking layers and provides
a $dE/dx$ resolution of $\sigma_{dE/dx} = 8.4\%$ for particle
identification and $\sigma_p/p = 1.7\% \sqrt{1+p^2}$ ($p$ in GeV) momentum
resolution for charged tracks. The barrel shower counter
(BSC), which covers $80\%$ of $4\pi$ solid angle, has an energy resolution
of $\sigma_E/E = 22\%/\sqrt{E}$ ($E$ in GeV) and a spatial resolution
of 7.9 mrad in $\phi$ and 2.3 cm in z, is located outside the TOF.
Outermost is a $\mu$ identification system, which consists of three
double layers of proportional tubes interspersed in the iron flux return
of the magnet.

\begin{figure}[htbp]
  \centerline{\hbox{ \hspace{0.2cm}
    \includegraphics[width=5.5cm]{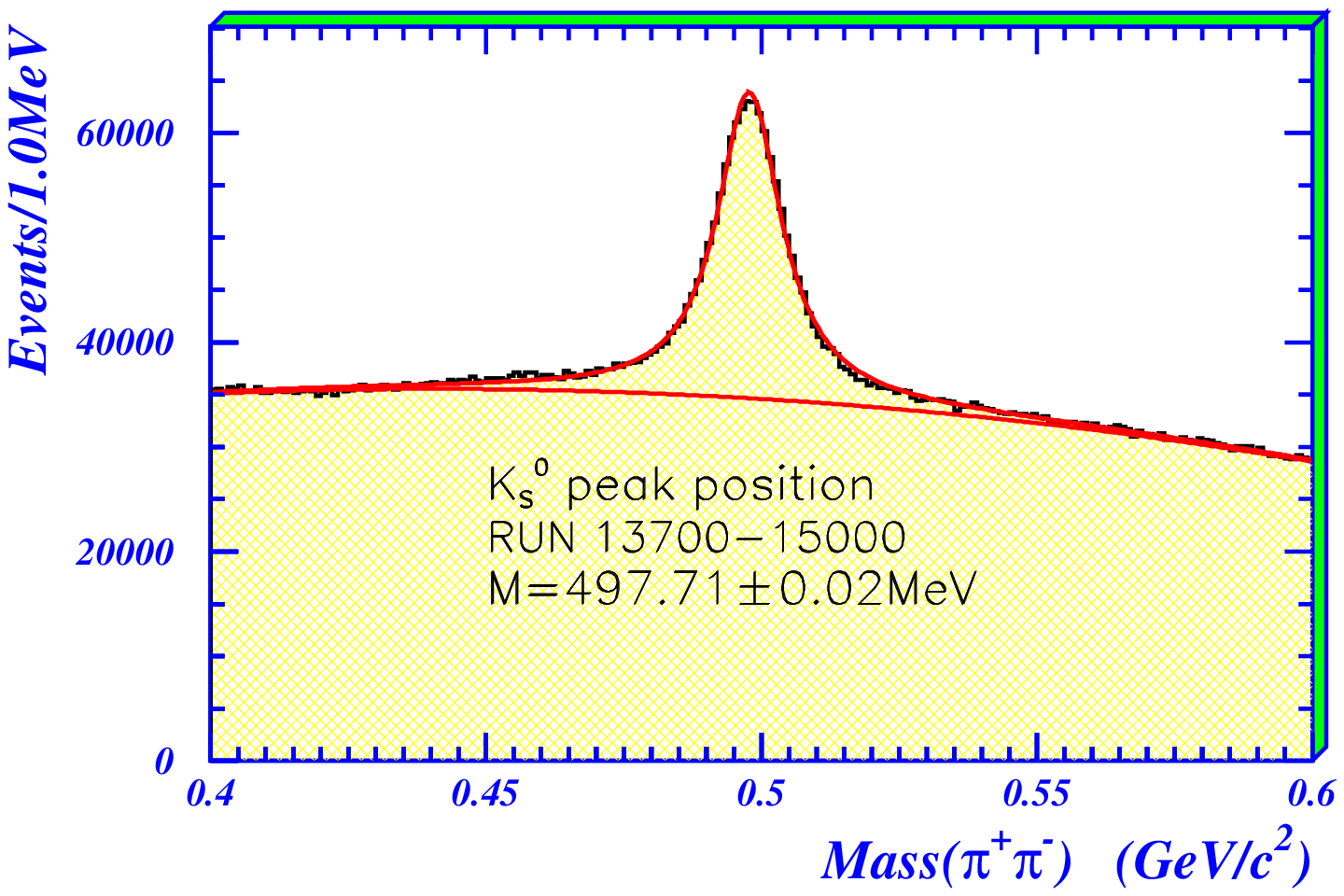}
    \hspace{0.3cm}
    \includegraphics[width=5.5cm]{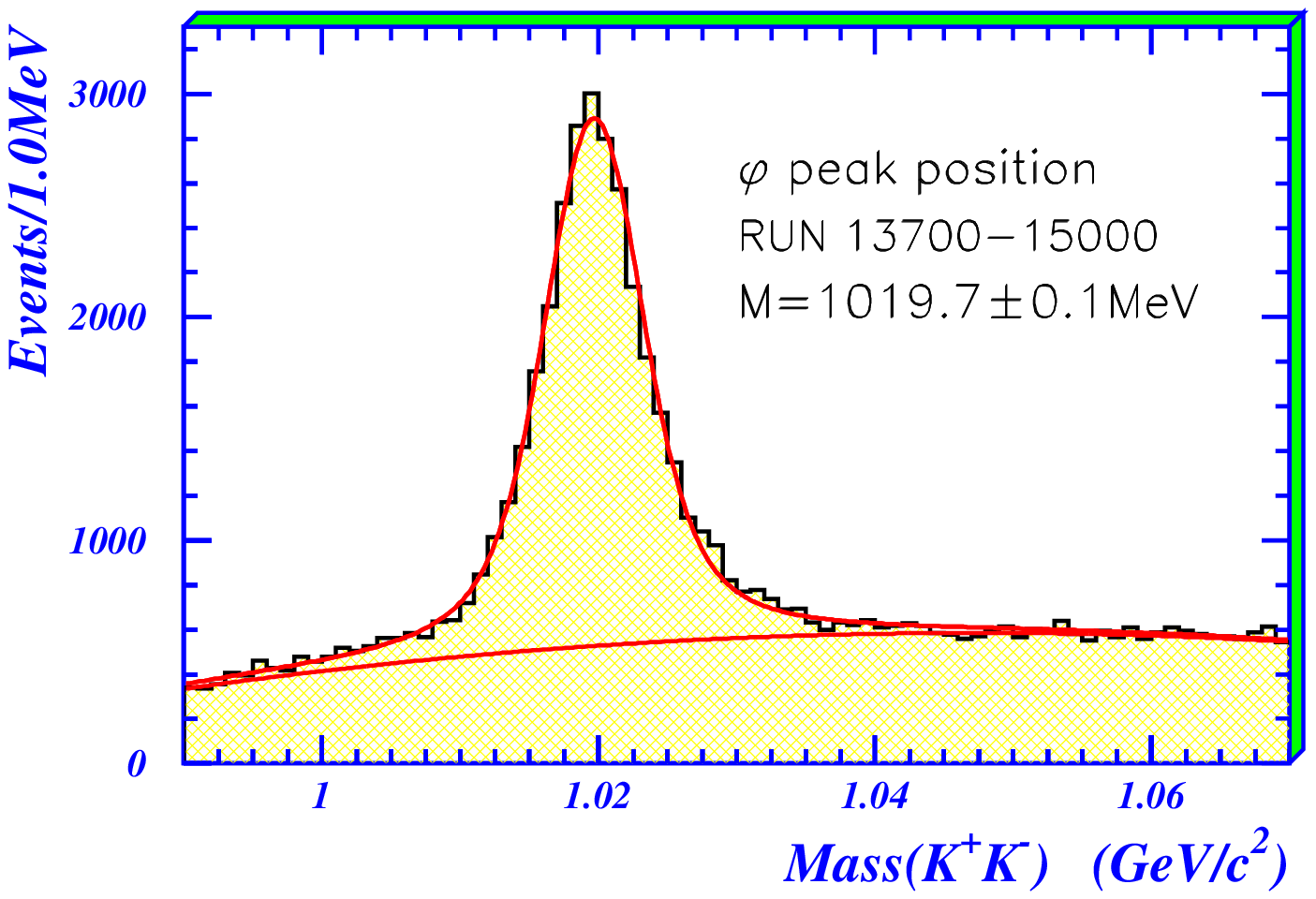}
    }
  }
 \caption{\it Inclusive $K_s^0$ and $\phi$ signals
    \label{inclusive}}
\end{figure}

With the upgraded BESII detector, till the summer
of 2001, about $5.8 \times 10^7 J/\psi$ events have been accumulated.
This is the largest $J/\psi$ data sample in the world. 
Fig. \ref{inclusive} shows the inclusive $K_s^0$ and $\phi$
signals, which indicates a good data quality and a large data sample.
\vspace{-0.5cm}

\subsection{Partial Wave Analysis of $J/\psi \to \gamma K \overline K$
(preliminary)}

\begin{figure}[htbp]
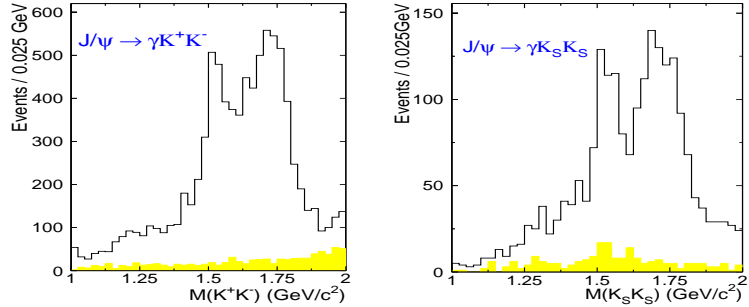

  \centerline{\hbox{ \hspace{0.2cm}
  \includegraphics[width=4.5cm,height=4.cm]{togkk2.0.epsi}
    \hspace{0.3cm}
  \includegraphics[width=4.7cm,height=4.cm]{kk-gks2.0.epsi}
    }
  }
 \caption{\it $K^+ K^-$ and $K_s^0 K_s^0$ invariant mass spectra
in $J/\psi \to \gamma K^+ K^-$ and $J/\psi \to \gamma K_s^0 K_s^0$
    \label{kkmass} }
\end{figure}

\begin{figure}[htbp]
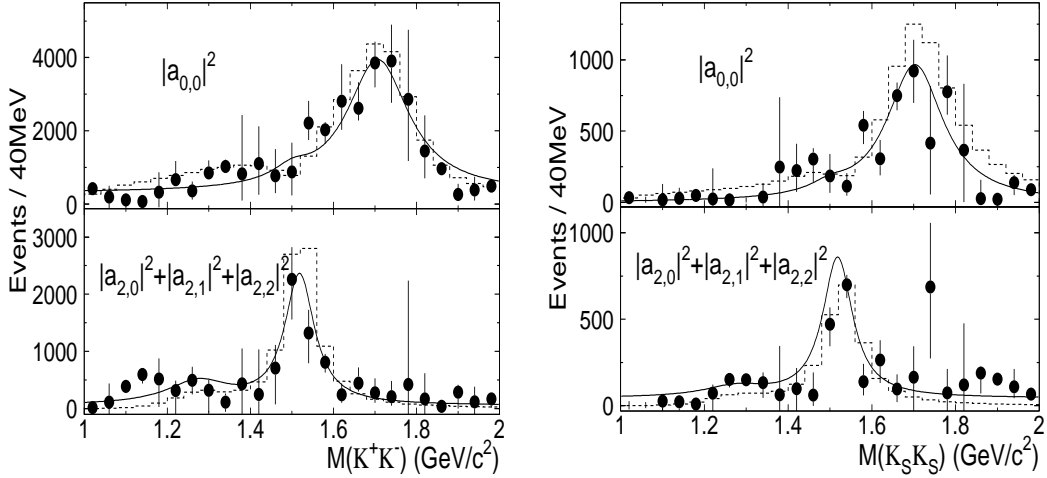

  \centerline{\hbox{ \hspace{0.2cm}
    \includegraphics[width=6.6cm,height=6.3cm]{heli2-kk.epsi}
    \hspace{0.3cm}
    \includegraphics[width=6.6cm,height=6.3cm]{heli2ks.epsi}
    }
  }
 \caption{\it PWA fit results of $K^+ K^-$ and $K_s^0 K_s^0$
    \label{kkfit} }
\end{figure}

The spin-parity of the structure around 1.7 GeV has a long history of 
uncertainty. 
Partial wave analyses (PWA) are applied to $J/\psi \to \gamma K^+ K^-$
and $\gamma K_s^0 K_s^0$ channels, based on BESII $5.8 \times 10^7 J/\psi$
events. Fig. \ref{kkmass} shows the invariant mass spectra of
$K^+ K^-$ and $K_s^0 K_s^0$, where $f_2'(1525)$ and the structure around 
1.7 GeV are clearly seen in both cases. 
\noindent
The amplitudes are fitted to the
relativistic covariant tensor expressions, and the maximum likelihood
method is applied in the fit. Global and slice fits are performed 
to $KK$ mass region of 1-2 GeV. For slice fit, 40MeV bin width is adopted.
Fig. \ref{kkfit} shows $0^{++}$ and
$2^{++}$ intensity distributions as a function of the invariant mass
of $K^+ K^-$ and $K_s^0 K_s^0$, where, dots with error bars are the
efficiency corrected data points, solid curves stand for the coherent 
superposition of the individual Breit-Wigner resonance fits and dashed 
histograms represent the global fit results.
Both $K^+ K^-$ and $K_s^0 K_s^0$ D wave intensity shows a clear $f_2'(1525)$
with the mass and width being $M=1518 \pm 6$ MeV and $\Gamma=84^{+28}_{-24}$
MeV. There is an evidence for $f_2(1270)$ and a weak $2^{++}$ intensity 
in the mass
region around 1.7 GeV. While $0^{++}$ is found to be dominant in 
1.7 GeV region, which is well determined by a Breit-Wigner resonance.
The mass and width of the $0^{++}$ component are:
$$M=1703^{+8}_{-10} MeV, ~~~\Gamma=163^{+27}_{-22}MeV$$

\subsection{Partial Wave Analysis of $J/\psi \to \gamma \pi^+ \pi^-$
(preliminary)}

\begin{figure}[htbp]
\centerline{\hbox{
\includegraphics[width=8cm,height=5.cm]{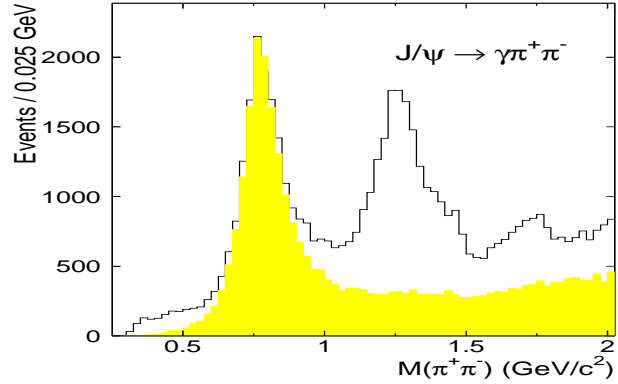}
       }}
 \caption{\it
     $\pi^+ \pi^-$ invariant mass spectrum in $J/\psi \to \gamma \pi^+ \pi^-$
    \label{pipimass} }
\end{figure}

\begin{figure}[htbp]
\centerline{\hbox{
\includegraphics[width=10.2cm,height=8.0cm,angle=-90]{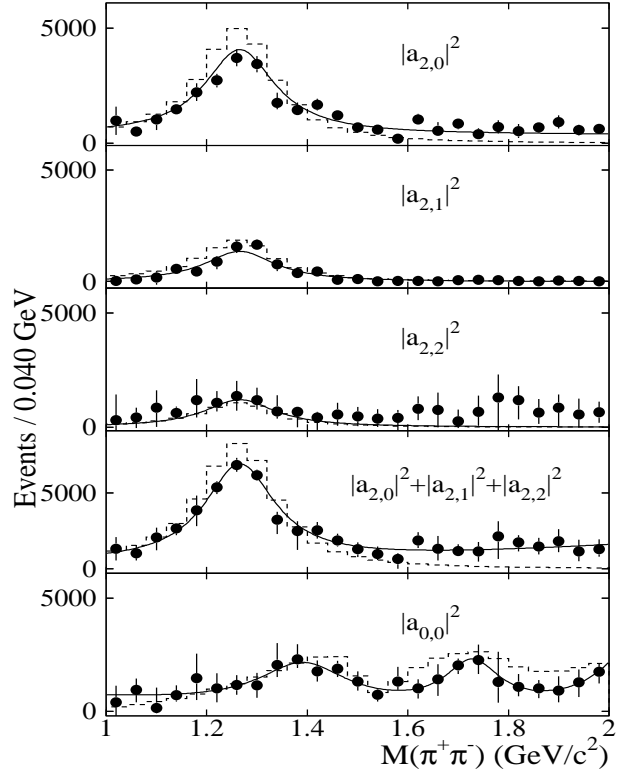}
       }}
 \caption{\it
    PWA fit results of $\pi^+ \pi^-$
    \label{pipifit} }
\end{figure}

Fig. \ref{pipimass} is the invariant mass spectrum of
$\pi^+ \pi^-$ in $J/\psi \to \gamma \pi^+ \pi^-$.
\noindent
Except for the well 
known $f_2(1270)$, a shoulder in the high mass region of $f_2(1270)$ 
and a bump near 1.7 GeV are seen. PWA is performed to $\pi^+ \pi^-$ 
mass in 1-2 GeV region. $0^{++}$ and $2^{++}$ intensity distributions 
are shown in Fig. \ref{pipifit}, where, the same as in $K\overline
K$ case, dots with error bars are the
efficiency corrected data points, solid curves stand for the coherent
superposition of the individual Breit-Wigner resonance fit and dashed
histograms represent the global fit results. 
\noindent
A strong $2^{++}$ is well
determined at around 1.27 GeV, and two $0^{++}$'s are observed at around 
1.4 and 1.7 GeV. If the $0^{++}$ at around 1.7 GeV is considered as 
the same one as in $K \overline K$ and the mass and width are fixed 
to be the values obtained from $K \overline K$ analyses, {\it i.e.} 
$$M=1703^{+8}_{-10} MeV, ~~\Gamma=163^{+27}_{-22}MeV,$$
the fit results give the masses and widths of $2^{++}$ and another 
$0^{++}$ as:
$$M=1266 \pm 6 MeV, ~~\Gamma=170^{+37}_{-30}MeV$$
$$M=1383^{+20}_{-18} MeV, ~~\Gamma=274^{+59}_{-128}MeV$$
The preliminary results indicate a relative $\pi \pi$ to $KK$ branching
ratio of around 30\%.

\subsection{Measurement of $\eta_c$ mass}

\begin{figure}[htbp]
  \centerline{\hbox{ \hspace{0.1cm}
 \includegraphics[width=4.8cm,height=3.6cm]{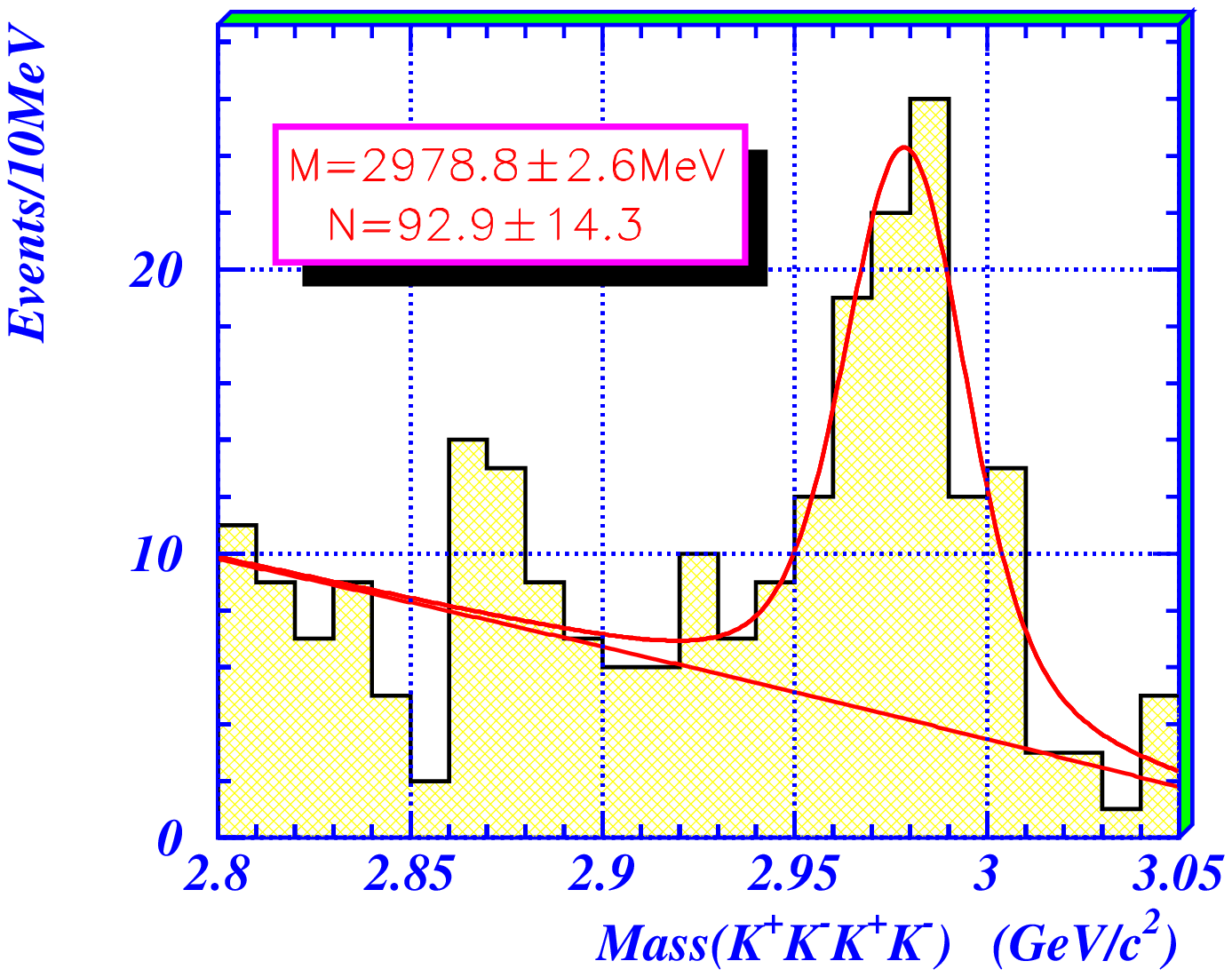}
    \hspace{0.1cm}
 \includegraphics[width=4.8cm,height=3.6cm]{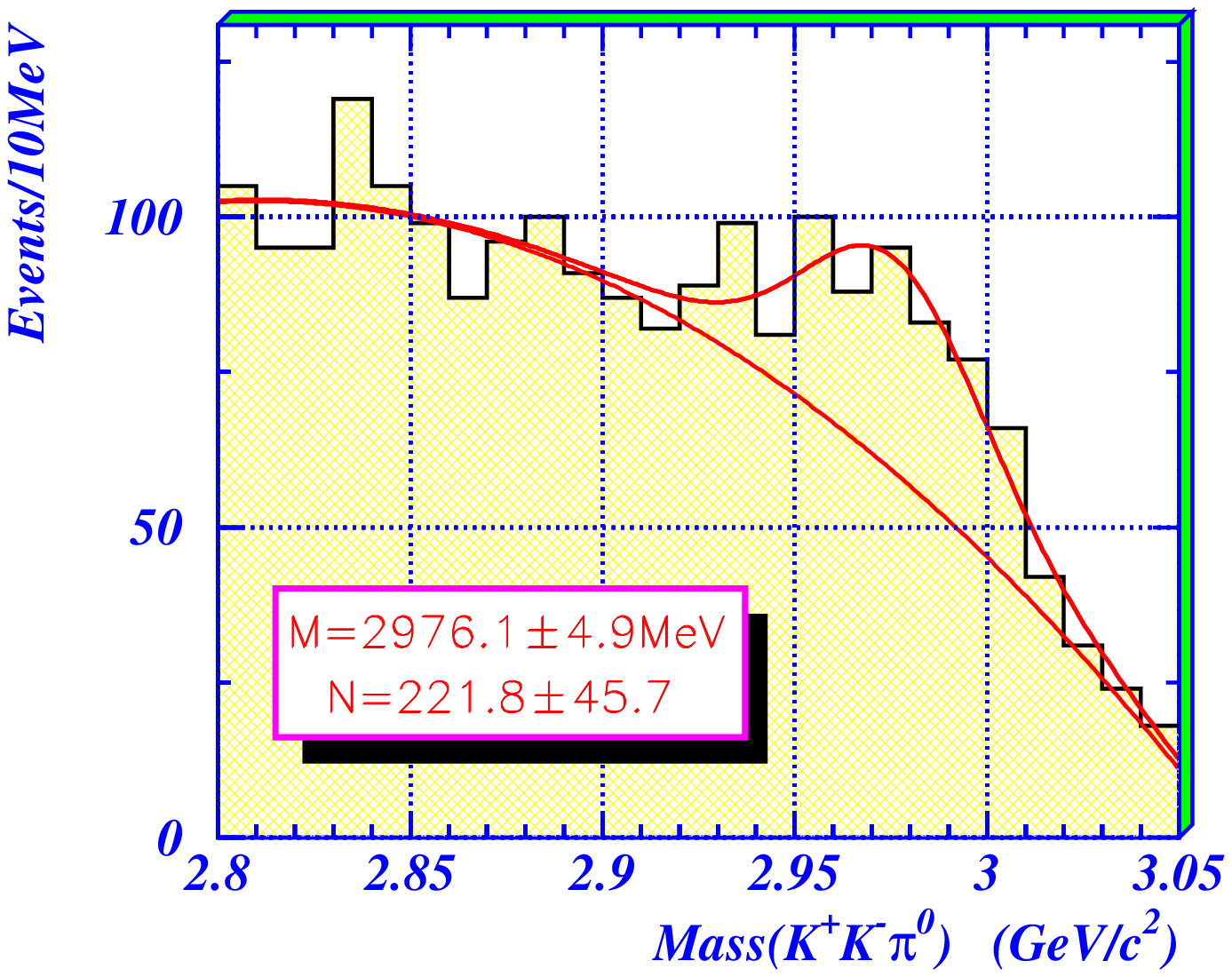}
    \hspace{0.1cm}
 \includegraphics[width=4.8cm,height=3.6cm]{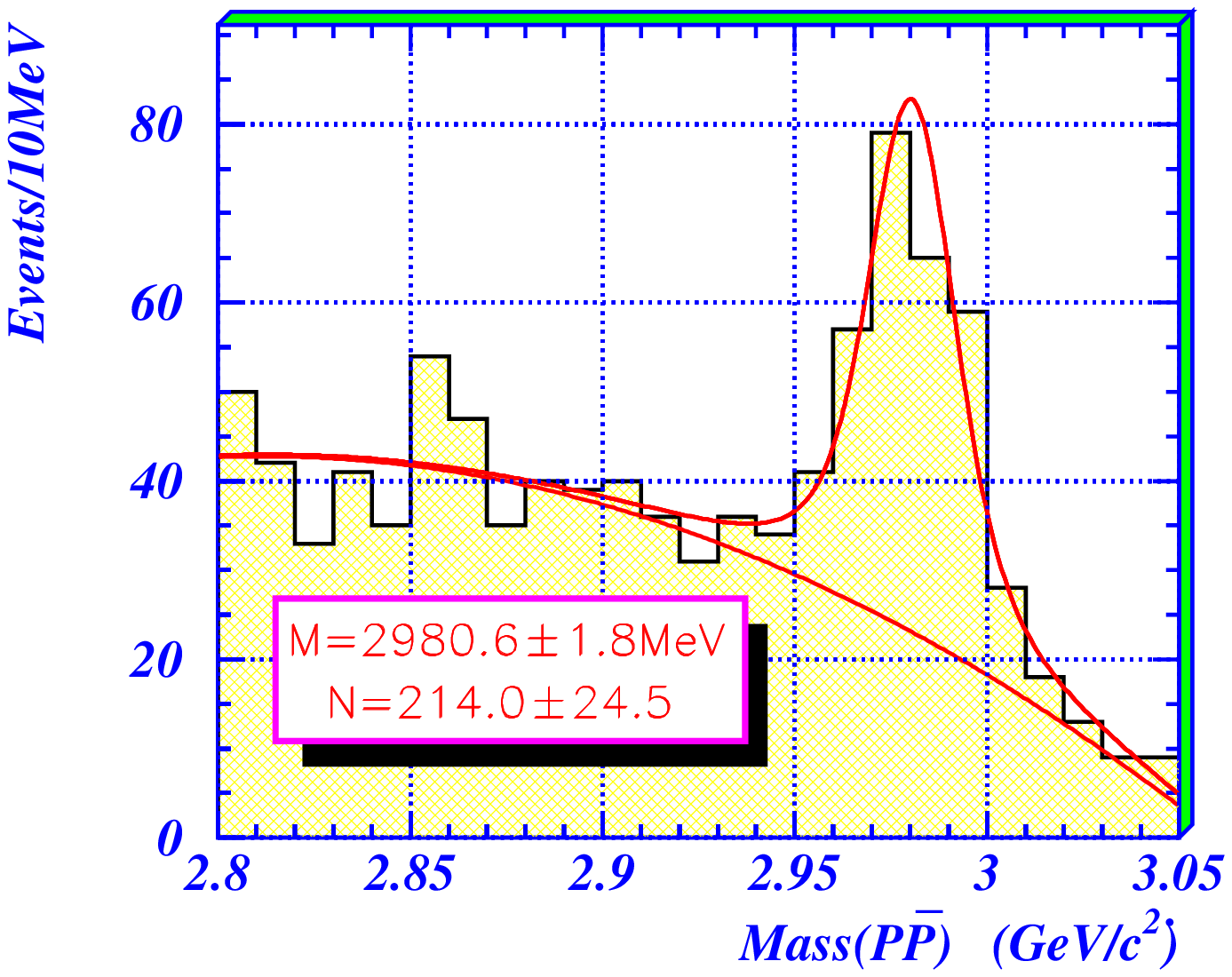}
    }}
 \caption{\it Mass spectra of $K^+K^-\pi^+\pi^-$,
$\pi^+\pi^-\pi^+\pi^-$, $K^\pm K_{S}^{0}\pi^\mp$,
$\phi\phi$, $K^+K^-\pi^0$, $p\bar{p}$
    \label{etac} }
\end{figure}

A precise knowledge of the mass difference between $J/\psi(1^{--})$ and 
$\eta_c(0^{-+})$ charmonium states is useful to the determination of the
strength of spin-spin interaction term in non-relativistic potential 
model. The mass of $J/\psi$ has been measured with 
high accuracy, while the $\eta_c$ mass measurements listed on PDG2000 
are very different and the fit to the measurements has a confidence 
level of only 0.001. Based on BESII $5.8 \times 10^7 J/\psi$ events, 
we measure the mass of $\eta_c$ from 6 channels: 
$J/\psi \to \gamma \eta_c$, $\eta_c \to K^+K^-\pi^+\pi^-$, 
$\pi^+\pi^-\pi^+\pi^-$, $K^\pm K_{S}^{0}\pi^\mp$,
$\phi\phi$, $K^+K^-\pi^0$ and $p\bar{p}$.
Fig. \ref{etac} shows $\eta_c$ signals in above 6 channels.
\begin{figure}[hp]
  \centerline{\hbox{ \hspace{0.1cm}
 \includegraphics[width=4.8cm,height=3.6cm]{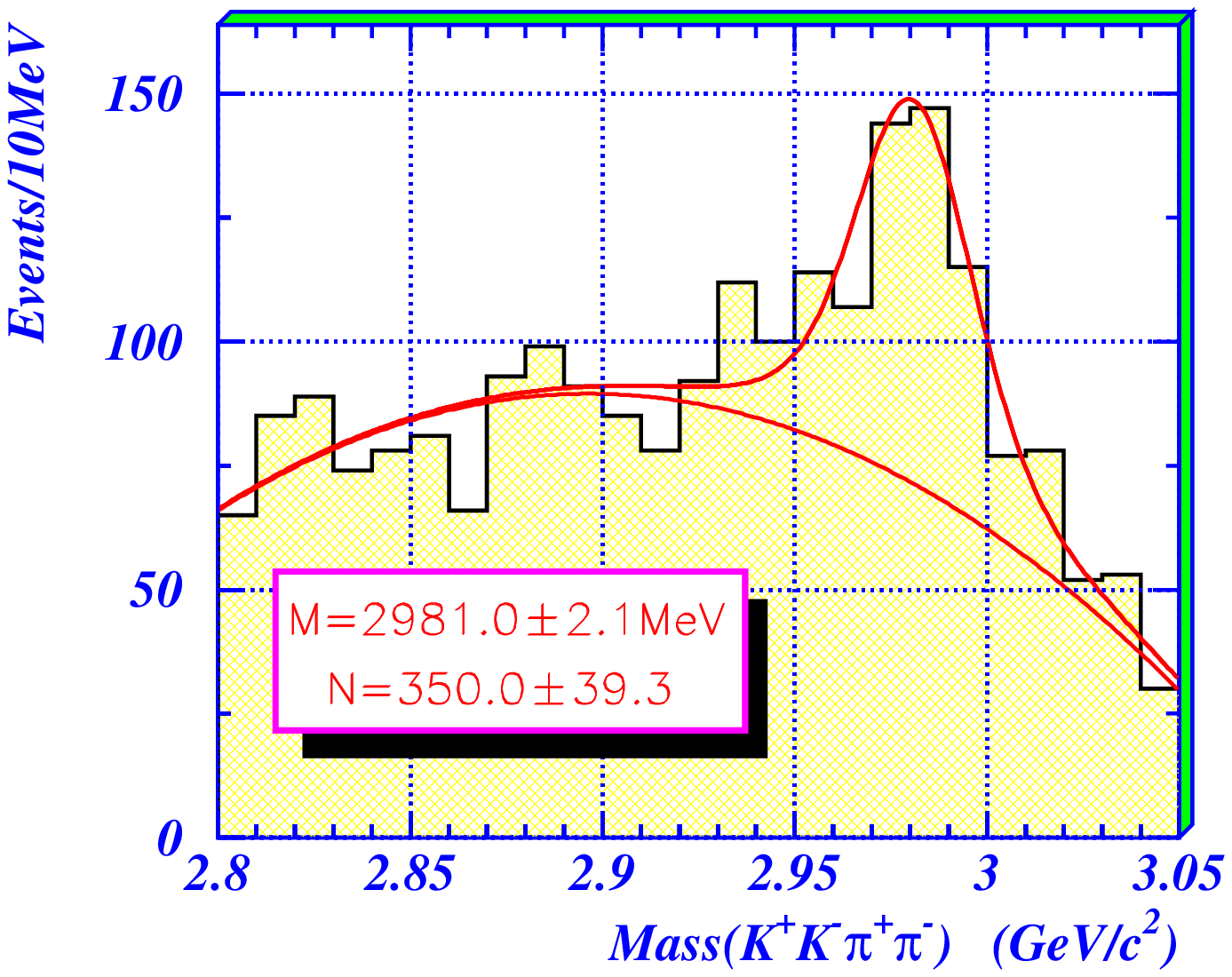}
    \hspace{0.1cm}
 \includegraphics[width=4.8cm,height=3.6cm]{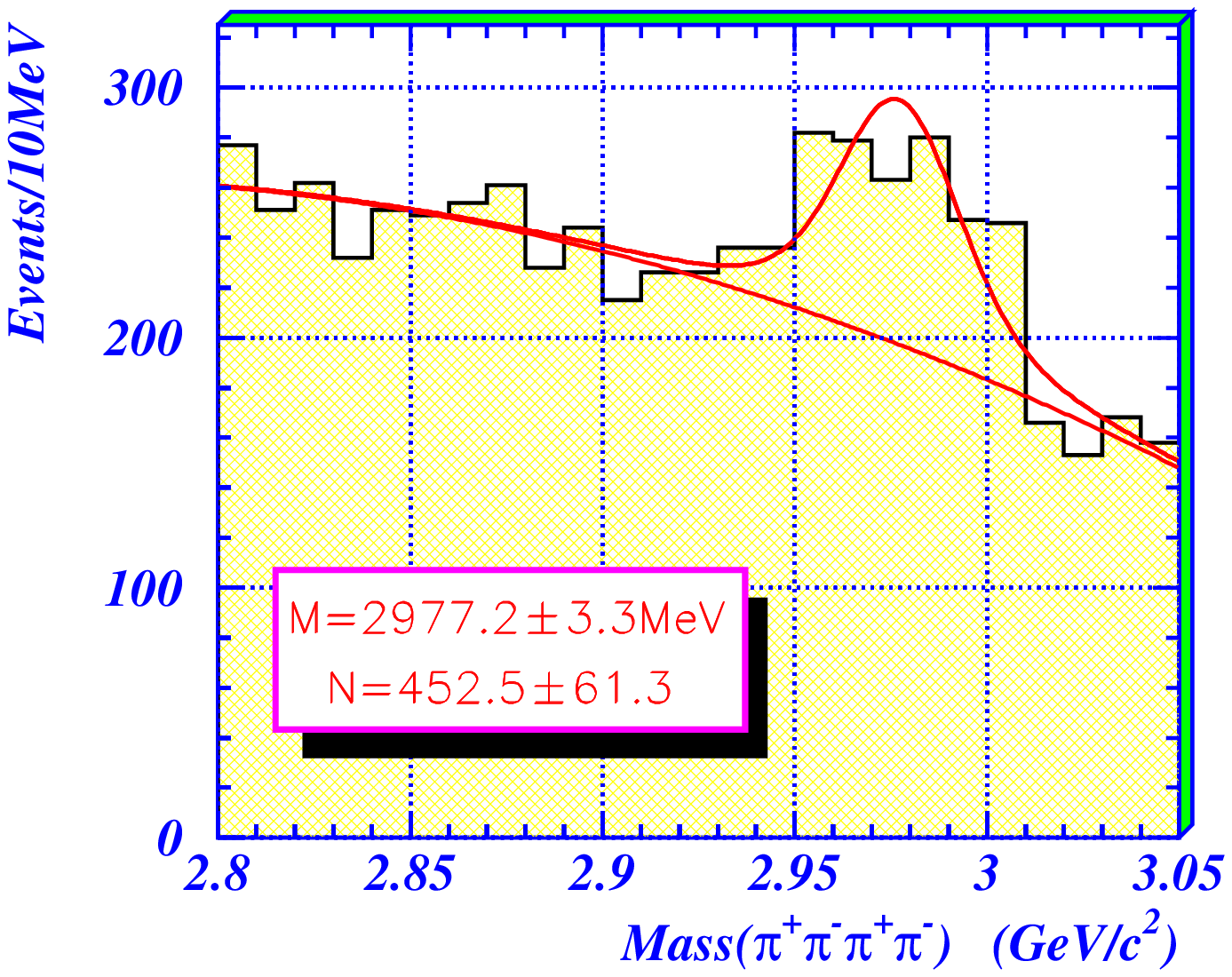}
    \hspace{0.1cm}
 \includegraphics[width=4.8cm,height=3.6cm]{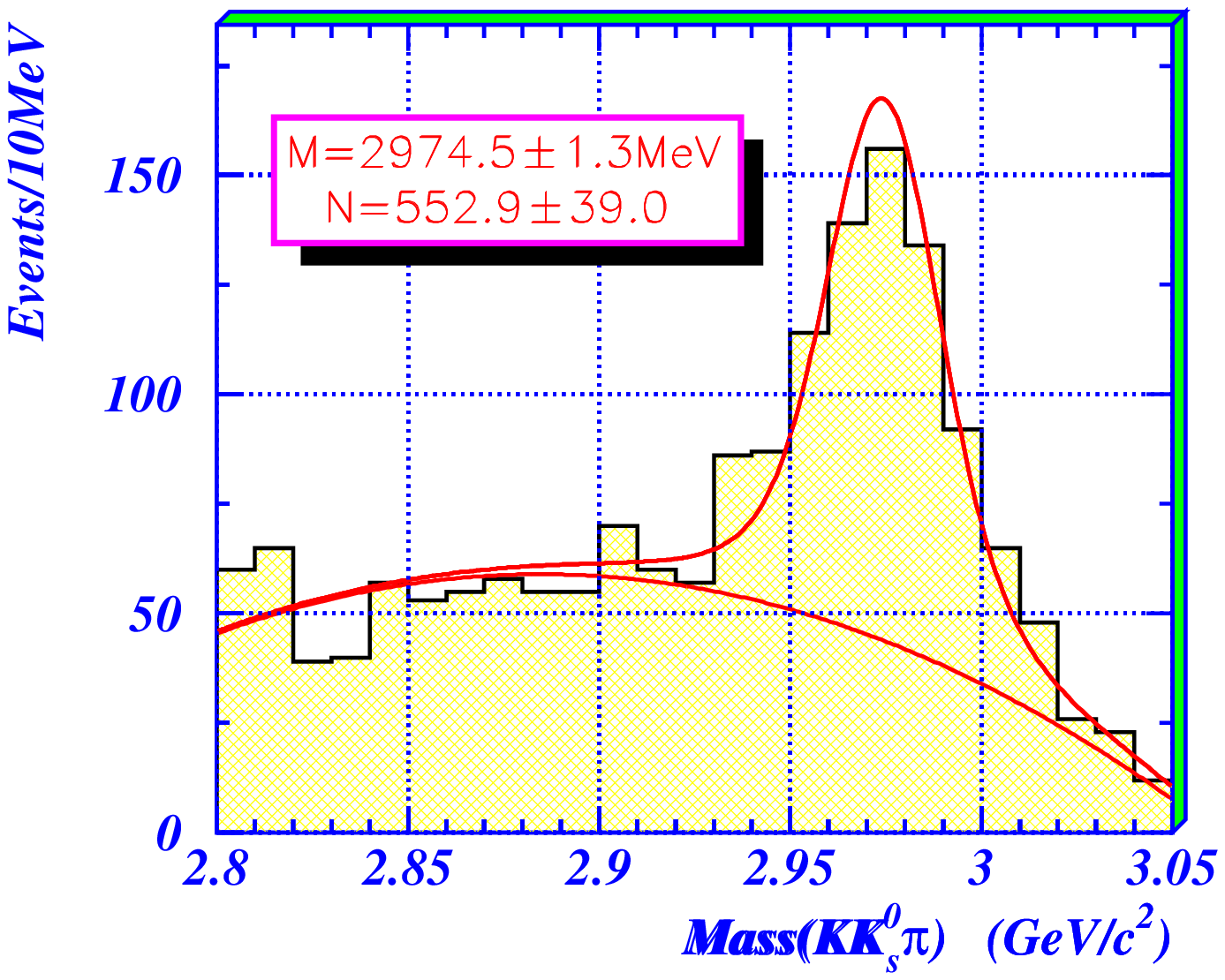}
    }
  }
\end{figure}
\vspace{-1.0cm}
\noindent
The fit values of the number of events and $\eta_c$ mass in the 
individual channels are listed in Table \ref{etac-tab} (the errors
are statistical only). In the fit, the width of $\eta_c$ is fixed at
16.5 MeV, which is the weighted average of PDG2000, 
BESI(2000)\cite{besietac} and recent CLEO two-photon collision experiment  
\cite{cleo-etac}. Combine the weighted average with the results from
the 5 channels, listed in Table \ref{etac-tab}, the mass of $\eta_c$
is given as $M_{\eta_c} = 2977.6 \pm 0.8$(stat.) MeV.

\begin{table}[htbp]
\centering
\caption{ \it Fit values of $\eta_c$ in the individual channels
}
\vskip 0.1 in
\begin{tabular}{|c|c|c|}\hline
Decays     & Number of events   &  $M_{\eta_c}$ (MeV)  \\\hline
$K^+K^-\pi^+\pi^-$      & $350.0 \pm 39.3$ & $2981.0 \pm 2.1$ \\\hline
$\pi^+\pi^-\pi^+\pi^-$  & $452.6 \pm 61.3$ & $2977.2 \pm 3.3$ \\\hline
$K^\pm K_{S}^{0}\pi^\mp$ & $552.9 \pm 39.0$ & $2974.5 \pm 1.3$ \\\hline
$\phi\phi$              & $92.9 \pm 14.3$ & $2978.8 \pm 2.6$ \\\hline
$p\bar{p}$       & $214.0 \pm 24.5$ & $2980.6 \pm 1.8$ \\\hline
\end{tabular}
\label{etac-tab}
\end{table}

\subsection{Near threshold structure in $J/\psi \to \gamma p \bar p$}

\begin{figure}[htbp]
  \centerline{\hbox{ \hspace{0.2cm}
  \includegraphics[width=8.cm,height=4.0cm]{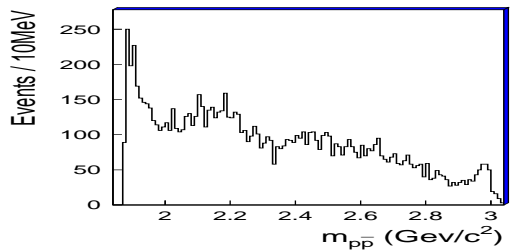}
    }}
\vspace{-0.5cm}
 \caption{\it $p \bar p$ mass spectrum
    \label{mpp} }
\end{figure}

\begin{figure}[htbp]
  \centerline{\hbox{ \hspace{0.2cm}
  \includegraphics[width=8.cm,height=6.0cm]{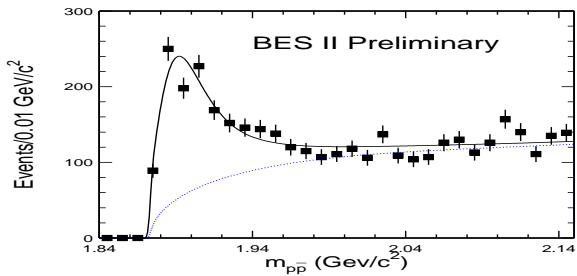}
    }
  }
 \caption{\it Fit to $p \bar p$ near threshold enhancement
    \label{gppfit} }
\end{figure}

There is an accumulation of evidence for anomalous behavior in the $p \bar p$
system near $2m_p$ mass threshold. A narrow dip-like structure at a
center of mass energy of $2m_p c^2$ was observed in $e^+ e^- \to 
hadrons$\cite{pp1}. A narrow dip structure which is just above $2m_p$ 
was also observed in diffractive photoproduction of $3\pi^+ 3\pi^-$ 
final states\cite{pp2}. Based on BESII $J/\psi$ data, we analyze 
$J/\psi \to \gamma p \bar p$. The invariant mass spectrum of $p \bar p$ 
is shown in Fig. \ref{mpp}. Except for $\eta_c$ 
signal, there is a clear enhancement near $2m_p$. 
An S-wave Breit-Wigner function, as shown below, is used to fit the
near threshold enhancement. 
$$ BW \sim \frac{M_0\Gamma_0 \*
(q/q_0)}{(M^2-M_0^2)^2+(M_0\Gamma_0 \*(q/q_0))^2}$$
Where, $q$ is the momentum of daughter particle, $q_0$ is the momentum
of daughter particle at peak. Fig. \ref{gppfit} shows the fit curve.
The preliminary fit gives (statistical errors only):
$$M=1894.1 \pm 1.4 MeV, ~~~\Gamma=48.0 \pm 5.9 MeV$$
\vspace{-2.cm}

\section{Summary}
$J/\psi$ radiative decay is a good laboratory in searches for glueball, 
the study of light hadron spectroscopy, as well as the glueball
spectroscopy. In this talk, we mainly reported the PWA results from 
$J/\psi \to \gamma K \overline K$, $\gamma \pi^+ \pi^-$, the measurement
of $\eta_c$ and the search for interesting states based on BESII
$5.8 \times 10^7$ $J/\psi$ data. More analyses are going on and more 
results will be produced very soon with this data set. CLEO is 
going to reduce the beam energy to collect a large amount of 
$J/\psi$ data in a few years. An experimental plan of upgrading 
BESII/BEPC to BESIII/BEPCII, with a luminosity of $10^{33}$ and 
a good detector is underway now. More results on the study of the 
light hadron spectroscopy and the search of the new form of hadronic 
states are expected in the near future.

\section{Acknowledgements}
We acknowledge the staff of the BEPC accelerator and IHEP computing
center for their efforts. The work was supported in part by the
National Natural Science Foundation of China under Contracts No.
19991480, No. 19825116 and No. 19605007, and by the Department of
Energy of US under Contracts No. DE-FG03-93ER40788 (Colorado State
University), No. DE-AC03-76SF00515 (SLAC), No. DE-FG03-94ER40833
(University of Hawaii) and No. DE-FG03-95ER40925 (University of
Texas at Dallas).

\end{document}